\documentstyle[11pt,paspconf,epsf,psfig,twoside]{article}

\def\ltsim{\raisebox{-.5ex}{$\;\stackrel{<}{\sim}\;$}}

\begin{document}

\noindent {\small To appear in {\it Magnetic Cataclysmic Variables}
Annapolis, Maryland. 13-17~Jul~1998.
A.S.P. Conference Series, Vol~???, p.~???.
eds: K.Mukai, C.Hellier. (San Francisco: ASP).}

\title{Disk-Anchored Magnetic Propellers -- A Cure for the SW~Sex Syndrome}

\author{Keith Horne}
\affil{Physics \&\ Astronomy, University of St.\ Andrews, KY16~9SS, 
Scotland, UK.  e-mail: kdh1@st-and.ac.uk}

\begin{abstract}

In AE~Aqr, magnetic fields transfer energy and angular momentum
from a rapidly spinning white dwarf to material in the gas stream
from the companion star, with the effect of spinning down the
white dwarf while flinging the gas stream material out of the binary
system.
This magnetic propeller
produces a host of observable signatures, chief among which are
broad single-peaked flaring emission lines with phase-shifted
orbital kinematics.
SW~Sex stars have accretion disks, but also broad single-peaked
phase-shifted emission lines similar to those seen in AE~Aqr.
We propose that a magnetic propeller similar to that which operates
in AE~Aqr is also at work in SW~Sex stars -- and to some extent
in all nova-like systems.
The propeller is anchored in the inner accretion disk, rather than
or in addition to the white dwarf.
Energy and angular momentum are thereby extracted from the inner disk
and transferred to gas stream material flowing above the disk,
which is consequently pitched out of the system.
This provides a non-local dissipationless angular momentum extraction
mechanism, which should result in cool inner disks with
temperature profiles flatter than $T\propto R^{-3/4}$,
as observed in eclipse mapping studies of nova-like variables.
The disk-anchored magnetic propeller model appears to explain
qualitatively most if not all of the peculiar features
of the SW~Sex syndrome.

\end{abstract}


\section{Introduction}

We now possess a fairly satisfactory understanding of the 
double-peaked emission lines from the accretion disks of
dwarf novae in quiescence.
Their twin peaks are the result of supersonic Doppler shifts
arising from gas in the accretion disk moving toward us on one side
and away from us on the opposite side of the disk
(Smak 1969; Horne \&\ Marsh 1986).
This picture is confirmed in eclipsing systems, 
where eclipses of the blue-shifted peak occur earlier than those
of the red-shifted peak.
The shapes of the Doppler profile wings indicate that
emission line surface brightnesses decrease as $R^{-3/2}
\propto \Omega_{\sc Kep}$, suggesting that the quiescent disk
emission lines are powered by magnetic activity 
similar to that which powers the chromospheres of rotating stars,
which scale as $\Omega_{\sc rot}$ (Horne \&\ Saar 1991).

There are, however, a large number of higher accretion rate systems
in which the emission lines display broad single-peaked profiles.
These include nearly edge-on systems whose eclipses indicate
that optically thick disks are present.
It is hard to understand how broad single-peaked lines can arise in
the disk flows in these systems.
The emission lines also have anomalous orbital kinematics --
radial velocity curves significantly delayed relative
to the white dwarf orbit.
A common interpretation of the phase-shifted velocity curves
is that the disk emission lines are affected by a broad S-wave
component related to the impact of the gas stream.
However, we will argue in this paper that the anomalous emission lines
arise instead from shocks in a broad equatorial fan of gas
that is being ejected from the system by
a magnetic propeller anchored in the inner accretion disk.

This disk-anchored magnetic propeller model is motivated by
recent breakthroughs in understanding the enigmatic system AE~Aqr,
in which a rapidly spinning white dwarf magnetosphere expels
the gas stream out of the system before an accretion disk
can form (Wynn, King \&\ Horne 1997).
It has recently been realized that an internal shock zone
should form in the exit stream to produce violently flaring broad
single-peaked emission lines at just the right place to account
for the anomalous orbital kinematics 
(Welsh, Horne \&\ Gomer 1998).

Encouraged by this success in understanding AE~Aqr,
we propose to extend the idea to disk-anchored
magnetic propellers operating in all high accretion rate disk systems.
These systems are afflicted by the ``SW~Sex syndrome'',
a cluster of anomalies including single-peaked emission lines
with skewed kinematics,
V-shaped eclipses implying flat temperature-radius profiles,
shallow offset line eclipses,
and narrow low-ionization absorption lines at phase 0.5.
Magnetic fields anchored in the Keplerian disk sweep forward
and apply a boost that expels gas stream material flowing above
the disk plane.
This working hypothesis offers a framework on which we can hang
all the SW~Sex anomalies.
The lesson for theorists is that magnetic links appear
to be transporting energy and angular momentum from the inner disk
to distant parts of the flow without associated viscous heating in the disk.

\section*{The SW~Sex Syndrome}

SW~Sex is the prototype of a sub-class of nova-like
(high accretion rate) cataclysmic variables (CVs)
that display a range of peculiarities that do not seem to fit
with the standard model of a cataclysmic variable
in which a gas stream from the companion star feeds an
accretion disk around a white dwarf.
The same cluster of anomalous phenomena is in fact seen 
to a greater or lesser extent
in most
if not all CVs with high accretion rates.
The SW~Sex anomalies may be summarized as follows:

1) Anomalous emission-line kinematics
	(Young, Schneider \&\ Shectman 1981; 
        Still, Dhillon \&\ Jones 1995).
        The emission lines have broad single-peaked profiles, even in
        eclipsing systems where lines from a Keplerian disk would
        produce two peaks.
        The radial velocity curves lag $30^\circ$ to $70^\circ$
        behind the orbit of the white dwarf.
        The emission is centred at low velocities in the lower-left
        quadrant of the Doppler map, a location corresponding
        to no part of the binary system.

2) Anomalous emission-line eclipses 
	(Young et al. 1981; Still et al. 1995).
        The emission-line eclipses are shallow and early compared
        with continuum eclipses.
        Red-shifted emission remains visible at mid eclipse.
        There is evidence for slow prograde rotation,
        eclipses being early on the blue side and late on the red side
        of the line profile.

3) Low-ionization absorption lines (Balmer, O{\sc I} $\lambda7773$),
        strongest around phase 0.5 when the secondary star is behind the disk
        (Young et al. 1981; Szkody \&\ Piche 1990; Smith et al. 1993).
        The absorptions have widths and blue-shifts of a
        few hundred km~s$^{-1}$.

4) V-shaped continuum eclipses imply a temperature deficit in the inner disk,
        giving $T(R)$ profiles flatter than the $T \propto R^{-3/4}$ profile
        of steady-state viscous disks 
	(Rutten, van Paradijs \&\ Tinbergen 1992).

Most of these anomalies were first described in LX~Ser by
Young et al. (1981), and by 1992 they were enshrined as 
the defining characteristics a sub-class
of CVs named for the prototype SW~Sex 
(Thorstensen et al. 1992).
The original classification of ``SW~Sex stars'' required eclipses
and 3h$<P_{orb}<4$h,
but longer period systems like BT~Mon ($P_{orb}=8$h) now qualify
(Smith, Dhillon \&\ Marsh 1998),
and systems like WX~Ari showing the emission-line anomalies but lacking
absorptions and/or eclipses are probably low-inclination cases
(Hellier, Ringwald \&\ Robinson 1994).
While particularly strong in systems with 3h$<P_{orb}<4$h,
this ``SW~Sex syndrome'' can be recognized in most
if not all CVs in high accretion states.

The SW~Sex stars have evaded satisfactory explanation for over 15 years.
Proposals include accretion disk winds 
(Honeycutt, Schlegel \&\ Kaitchuck 1986; 
Murray \&\ Chiang 1996; 1997),
magnetic white dwarfs disrupting the inner disk
(Williams 1989; Wood, Abbott \&\ Shafter 1992),
and gas streams overflowing the disk surface 
(Shafter, Hessman \&\ Zhang 1988; Szkody \&\ Piche~1990; 
Hellier \&\ Robinson 1994).
The gas stream overflow model has been developed to the stage
of predicting trailed spectrograms that do bear resemblance to those
observed (Hellier 1998).
While each idea explains part of the phenomenology, none seems
to be entirely satisfactory.
The syndrome is so widespread that we must admit that
some important element is missing in our standard picture
of the accretion flows in CVs.

\section*{The Magnetic Propeller in AE~Aqr}

When facing a difficult puzzle, one helpful strategy 
is to examine extreme cases in search of clues.
AE~Aqr is an extreme CV in many respects.
For example, it is detected over a spectacular range of energies,
from radio 
(Bastian, Dulk \&\ Chanmugam 1988; Abada-Simon et al. 1993)
to perhaps TeV gamma rays (Meintjes 1994).
Among its problematic behaviours
are broad single-peaked emission lines that lag behind
the white dwarf orbit by some $70^\circ$
(Welsh, Horne \&\ Gomer 1993)
-- a classic SW~Sex symptom.
We suggest that AE~Aqr is the key to understanding the SW~Sex syndrome.

All CVs flicker,
with typical amplitudes of 5-30\% (Bruch 1992),
but AE~Aqr flickers spectacularly (Patterson 1979; 
van Paradijs, Kraakman \&\ van Amerongen 1989;
Bruch 1991; Eracleous \&\ Horne 1996).
Its line and continuum fluxes can vary by factors of 2 to 3,
with 10-min rise and decline times.
Transitions between quiet and active states occur on timescales of hours.
{\it HST} spectra reveal nearly synchronous flaring in the continuum and in
a host of high and low ionization permitted and semi-forbidden emission
lines (Eracleous \&\ Horne 1996).
The wide ionization and density ranges suggest that shocks power the lines.
All the emission lines share the anomalous orbital kinematics.

AE~Aqr harbours the fastest-rotating magnetic white dwarf.
Coherent X-ray, UV, and optical oscillations
indicate that the spin period is $P_{spin}=33$s,
with 2 peaks per cycle arising from hot spots on opposite sides of
the white dwarf, suggesting accretion onto opposite magnetic poles.
The oscillation amplitudes are $0.1-1$\% in the optical (Patterson 1979),
but rise to 40\% in the UV (Eracleous et al. 1994).
Remarkably, the UV oscillation amplitude is independent of the flaring state,
indicating that the flares are not accretion events.
{\it HST} pulse timing accurately traces the white dwarf orbit.
Optical pulse timing over a 13-year baseline shows that $P_{spin}$
is increasing with $P/\dot{P} \sim 2\times10^7$yr
(de Jager et al. 1994).
Rotational energy extraction from the white dwarf
at $I \omega \dot{\omega} \sim 6\times10^{33}$~erg~s$^{-1}$
exceeds the observed luminosity
$\nu L_\nu \sim 10^{32}$~erg~s$^{-1}$, thus
AE~Aqr can be powered by rotation rather than accretion.

What happens when the gas stream encounters this rapidly
spinning magnetosphere?
In AM~Her stars ($P_{spin}=P_{orb}$), the gas stream is gradually
stripped as material becomes threadded onto and slides
down along field lines to accretion shocks near the magnetic poles.
In AE~Aqr ($P_{spin}<<P_{orb}$), a drizzle of magnetic accretion
produces the white dwarf hot spots that
give rise to the spin pulsations,
but the magnetosphere rotates so rapidly that much of the threadded gas
is flung outward to the crests of magnetic loops.
The light cylinder, where co-rotation requires light speed,
is comparable in size to the separation of the binary system.
Particles trapped in the rapidly rotating magnetosphere,
on field lines that are shaken each time they sweep past
the companion star and gas stream,
get pumped up to relativistic velocities,
accounting for the observed radio synchrotron emission,
and perhaps the TeV gamma rays (Kuijpers et al. 1997).

\begin{figure*}
\begin{picture}(0,0)(20,30)
\put(0,0){\includegraphics{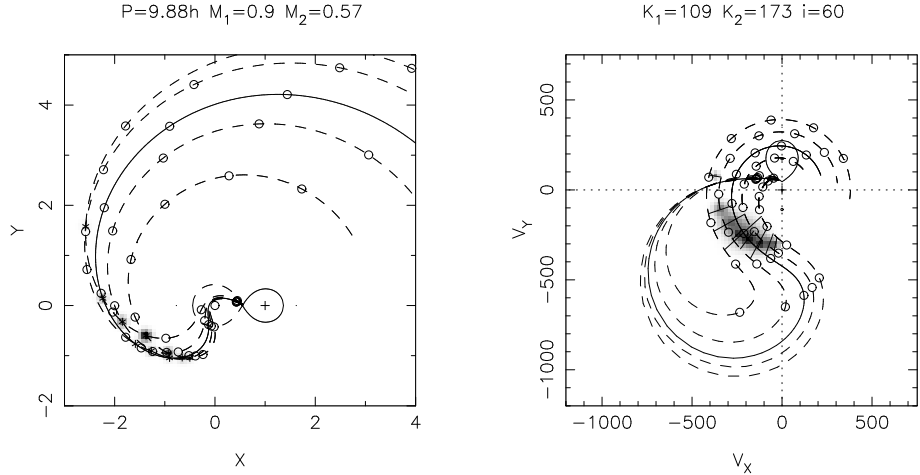}}
\noindent \end{picture}
\vspace{65mm}
\caption{Trajectories of diamagnetic blobs interacting with
a rapidly spinning white dwarf magnetosphere in AE~Aqr.
The trajectories are shown in both position and Doppler coordinates.
Open circles mark time of flight in units of $0.1~P_{\sc orb}$.
Asterisks and grey Gaussian patches
mark locations in the exit stream where
fluffy blobs overtake and collide with compact blobs,
producing fireballs.
This occurs at low velocities in the lower-left quadrant of the
Doppler map, at the location corresponding to the observed emission lines.
\label{fig_magprop}
}
\end{figure*}

Rapid rotation has another important effect:
high ram pressure in the rotating frame of the magnetosphere
reduces the stripping rate,
allowing the flow to remain largely dia-magnetic,
slipping between the rotating magnetic field lines.
So long as ram pressure exceeds the external magnetic pressure,
the flow traces a roughly ballistic trajectory while gradually
being dragged forward toward co-rotation with the magnetosphere
(King 1993; Wynn \&\ King 1995).
In AE~Aqr, the rapidly spinning magnetosphere acts as a ``magnetic propeller'',
boosting gas to escape velocity and ejecting it out of the binary system
(Wynn, King \&\ Horne 1997; left-hand side of Fig.~\ref{fig_magprop}).

On a Doppler map (right-hand side of Fig.~\ref{fig_magprop}),
the flow of dia-magnetic blobs initially traces a ballistic trajectory,
moving leftward (toward negative $V_X$) from the L1 point.
The blobs then circulate counter-clockwise around a high-velocity loop
in the lower-left quadrant, reaching a maximum velocity of
$\sim 1000$~km~s$^{-1}$ as they fly by the white dwarf.
Finally, the blobs decelerate to a terminal velocity 
of $200-500$~km~s$^{-1}$, depending on blob properties,
as they climb out of the potential well.
Here the trajectories pass right through the region required to
account for the peculiar kinematics of AE~Aqr's emission lines.
After escaping from the binary, blobs coast outward at constant velocity,
and hence circulate clockwise around the origin of the Doppler map
at their terminal velocities.  This is because the Doppler map
is in the rotating frame of the binary.

The exit stream of the magnetic propeller model places 
gas at low-velocity in the lower-left quadrant of the Doppler map,
as required to account for the orbital kinematics of AE~Aqr's
emission lines.
No part of the binary system moves with this velocity.
While this is a great success, it remains a puzzle
why the flaring line and continuum
emission do not arise near closest approach,
where interaction with the magnetosphere is strongest,
but rather several hours later
in the decelerating exit stream somewhat before
reaching the terminal velocity (Fig.~\ref{fig_magprop}).

\section*{The Flare Mechanism in AE~Aqr}

Why does the magnetic propeller flow produce wildly flaring emission
lines, and why do they occur on the way out rather than at closest approach?
The first step toward answering this question is to note that the
trajectories calculated for diamagnetic blobs of various sizes and densities
indicate that the magnetic propeller acts as a ``blob spectrometer''.
Small dense blobs with low drag coefficients
penetrate more deeply into the magnetosphere
and emerge at larger azimuths than more ``fluffy'' blobs that
are more readily repelled by the spinning magnetosphere.
Thus a train of incoming blobs diverse in size and density
is segretated in the magnetosphere to emerge as a broad fan
sorted by the effective drag coefficient.
The second step is to note that fluffy blobs reach a higher terminal
velocity, and can therefore overtake and collide with compact blobs that
passed closest approach at an earlier time.
The crossing points of blob trajectories (Fig~\ref{fig_magprop})
indicate that blob-blob collisions occur in an arc-shaped zone
of the exit stream well outside the Roche lobe.
Each blob-blob collision spawns a hot expanding fireball
representing a single flare.
If the blob rate is low, most blobs pass quietly through the system
without suffering a major collision (quiet states),
but above some threshold blob-blob collisions become a frequent
occurrence (flaring states).
Velocity vectors in the blob-blob collision zone
nicely match the observed emission-line kinematics
(Welsh et al. 1998).
Velocity differences between colliding blobs
are $\Delta V \ltsim 300$~km~s$^{-1}$, thus the fireballs
may initially reach temperatures up to $\sim 10^7$K.
Rapidly expanding and adiabatically cooling fireballs emit
a flare of continuum and line emission, with decreasing density
and ionization.
The situation may be not unlike a supernova explosion, albeit with
lower velocity, smaller size, and shorter timescale.
In this way the magnetic propeller model provides a natural mechanism
to explain the wild flaring behaviour and skewed orbital kinematics
of the emission lines in AE~Aqr.

It seems that AE~Aqr's rapidly spinning magnetized white dwarf
prevents formation of a disk by driving a rotation-powered outflow.
Energy and angular momentum are extracted from the white dwarf,
conveyed by magnetic field lines to the gas stream,
and expelled from the system.
Remarkably, the observations show that the magnetic boost
is a gentle one, producing
little dissipation in the region of closest approach.
Most if not all of AE~Aqr's exotic behaviour
can be understood in the framework of this
magnetic propeller model.

The processes we observe in AE~Aqr and other magnetic
cataclysmic variables should let us understand magnetic propellers
well enough to identify their observational signatures
and physical consequences in other types of systems.
Magnetic propellers may represent a generic non-local dissipationless
angular momentum extraction mechanism that could have a huge impact
on our understanding of accretion flows in general.
This prospect provides major motivation for studying magnetic
accretors in detail.

\section*{Disk-Anchored Magnetic Propellers}

The magnetic propeller in AE~Aqr offers a natural mechanism
that produces highly variable emission lines with broad single-peaked
velocity profiles and orbital kinematics lagging behind the white dwarf orbit.
We witness these same anomalies in the emission lines of
SW~Sex stars and in fact most nova-like (high accretion rate) CVs.
Could magnetic propellers be operating in these systems too?

Imagine tipping AE~Aqr over until the secondary star eclipses the
white dwarf.
We would then see a shallow partial eclipse of the exit stream
occurring a bit earlier than the deep eclipse of the white dwarf.
Red-shifted emission would remain visible at mid eclipse.
Low-ionization blue-shifted absorption lines would be seen over a range of
phases around 0.5, when we view the white dwarf through the exit stream.
Thus it seems that most of the SW~Sex anomalies are reproduced
at least qualitatively by the magnetic propeller model.
We need only add a disk to produce deep continuum eclipses.

We therefore propose that magnetic propellers can be anchored in
accretion disks as well as white dwarfs.
The gas stream from the companion star deposits most of its
material when it crashes into the disk's rim,
but fringes of the gas stream clear the rim and flow inward
on ballistic trajectories above and below the disk surface (Lubow 1989).
This gas interacts with magnetic fields anchored in the disk below.
Closed loops near the disk surface drag the gas toward
the local Kepler speed, encouraging it to crash down on the disk surface.
But interactions higher above the disk plane
increasingly involve open field lines and large loops
anchored at smaller radii.
These magnetic structures, bending out and back as their footpoints
circle the inner disk, drag the gas forward, up, and out.
There may be a watershed below which the gas falls onto the disk surface
and above which the gas is boosted up to escape velocity 
and exits the system in a manner similar to the flow in AE~Aqr.
We propose that this general scenario accounts
for the anomalous emission-line behaviour
and the narrow absorptions at phase 0.5 in the SW~Sex stars.

Disk-anchored magnetic propellers launch an equatorial outflow
by extracting energy and angular momentum from the inner disk
where the field lines are anchored.
Removal of angular momentum from the inner disk
drives accretion without depositing heat.
This lowers inner disk temperatures below
the $T\propto R^{-3/4}$ law of steady-state viscous disks,
accounting for the flatter $T(R)$ profiles 
inferred from the V-shaped eclipses seen in SW~Sex stars.

Our present heuristic sketch of a model is strongly motivated by the 
similarity of the anomalous phenomena observed in AE~Aqr,
SW~Sex, and to some extent in all nova-like (high accretion rate) CVs.
This new synthesis seems quite promising, but needs to be tested
by the development of more detailed predictions and comparisons
with observations.
These tests include computing trailed spectrograms, Doppler maps, and
eclipse effects for comparison with observations of SW~Sex stars.
The time-dependent spectra of the fireballs should be computed
for comparison with the observed lightcurves and
ultraviolet emission-line spectrum of the flares in AE~Aqr.
Such tests will be a focus for future work.

The main implication for accretion theorists may be that
energy and angular momentum is being transported from disks
to outflows via long magnetic links
without associated dissipation in the disk.
The SW~Sex syndrome is the observational signature indicating
that this non-local transport mechanism is occurring in CVs.
Similar effects may be expected in AGN and protostellar disks,
whenever incoming gas flows above the disk surface.

There will certainly be consequences for dwarf nova outbursts
models, and for the long-term evolution of mass-transfer binary stars.
For example, consider the 2-3 hour gap in the orbital period distribution of
CVs.
Angular momentum losses due to gravitational radiation
and the companion star's magnetic wind drive CV evolution
toward shorter orbit periods.
Magnetic wind losses are thought to disappear at short orbital
periods ($P_{\sc orb}<3$h) becuase the dynamo generating the star's
magnetic field may shut down when the star, stripped down to 
$\sim 0.3~M_\odot$, becomes fully convective.
Mass transfer then ceases when the star loses contact with its Roche lobe
at $P_{\sc orb}\sim3$h,  and resumes when contact is re-established
at $P_{\sc orb}\sim2$h.
This is an attractive scenario, but available evidence 
shows no decrease in the magnetic activity of single stars
below this mass (e.g. Hawley, Gizis \&\ Reid 1997).
If CV evolution in long-period systems is driven instead by
angular momentum losses in magnetic propeller outflows,
the period gap might then arise from a cessation of this effect.
This might occur at a critical mass ratio, e.g. when the 3:1
or other resonances enter the disk, causing the outer rim to thicken
enough to prevent stream overflow.
Such speculations as these may be worthy of further investigation.

\acknowledgments
I am grateful for discussions on these topics
with many friends and collaborators including 
Vik Dhillon, Martin Still, Danny Steeghs, Coel Hellier, 
Tom Marsh, Graham Wynn, Andy King, and Mario Livio.


\begin{references}



\reference Abada-Simon, M., Lecacheux, A., Bastian, T.S.,
	Bookbinder, J.A., Dulk, G.A.
	1993, ApJ, 406, 692

\reference Bastian, T.S., Dulk, G.A., Chanmugam, G.
	1988, ApJ, 324, 431

\reference Bruch, A.
	1991 A\&A, 251, 59

\reference Bruch, A.
	1992, A\&A, 266, 237

\reference Eracleous, M., Horne, K.
	1996, ApJ, 471, 427

\reference Eracleous, M., Horne, K., Robinson, E.L.,
	Zhange, E.-H., Marsh, T.R., Wood, J
	1994, ApJ, 433, 313

\reference Hawley, S.L., Gizis, J.E., Reid, N.I.
	1997, AJ, 113, 1458

\reference Hellier, C.,
	1998, PASP, 746, 420

\reference Hellier, C., Ringwald, F.A., Robinson, E.L.
	1994, A\&A, 289, 148

\reference Hellier, C., Robinson, E.L.
	1994, ApJ, 431, L107

\reference Honeycutt, R.K., Schlegel, E.M., Kaitchuck, R.H.
	1986, ApJ 302, 388

\reference Horne, K., Marsh, T.R.
	1986, MNRAS, 218, 761

\reference Horne, K., Saar, S.H.
	1991, ApJ, 374, L55

\reference de Jager, O.C., Meintjes, P.J., O'Donoghue, D., Robinson, E.L.
	1994, MNRAS 267, 577

\reference King, A.R.
	1993, MNRAS, 261, 144

\reference Kuijpers, J., Fletcher, L., Abada-Simon, M., 
	Horne, K., Raadu, M.A., Ramsay, G., Steeghs, D.
	1997, A\&A, 322, 242

\reference Lubow, S.
	1989, ApJ, 340, 1064

\reference Meintjes, P.J. et al.
	1994, ApJ, 434, 292

\reference Patterson, J.
	1979, ApJ, 234, 978

\reference van Paradijs, J., Kraakman, H., van Amerongen, S.
	1989, A\&A Suppl. 79, 205

\reference Rutten, R.G.M., van Paradijs, J., Tinbergen, J.
	1992, A\&A, 260, 213

\reference Smak, J.
	1969, Acta Astr., 19, 155

\reference Smith, R.C., Fiddik, R.J., Hawkins, N.A., Catalan, M.S.
	1993, MNRAS, 264, 619

\reference Smith, D.A., Dhillon, V.S., Marsh, T.R.
	1998, MNRAS, 296, 465

\reference Still, M.D., Dhillon, V.S., Jones, D.H.P.
	1995, MNRAS, 273, 863

\reference Szkody, P., Piche, F.
	1990, ApJ, 361, 235

\reference Thorstensen, J.R., Ringwald, F.A., Wade, R.A.,
	Schmidt, G.D., Norsworthy, J.E.
	1992, AJ, 102, 272

\reference Welsh, W.F., Horne, K., Gomer, R.
	1993, ApJ, 410, L39

\reference Welsh, W.F., Horne, K., Gomer, R.
	1998, MNRAS, 298, 285

\reference Williams, R.E.
	1989, AJ, 97, 1752

\reference Wood, J.H., Abbott, T.M.C., Shafter, A.W.
	1992, ApJ 393, 729

\reference Wynn, G.A., King, A.R.
	1995, MNRAS, 275, 9

\reference Wynn, G.A., King, A.R., Horne, K.
	1997, MNRAS, 286, 436

\reference Young, P., Schneider, D.P., Shectman, S.A.
	1981, ApJ, 244, 259






































\end{references}
\end{document}